\documentstyle[sprocl]{article}

\input{axodraw.sty}
\def\Journal#1#2#3#4{{#1} {\bf #2}, #3 (#4)}

\def\NPB{{\em Nucl. Phys.} B}

\def\PRC{{\em Phys. Rev.} C}

\def\AP{\em Ann.Phys.(N.Y.)}

\def\be{\begin{equation}}
\def\ee{\end{equation}}
\def\bea{\begin{eqnarray}}
\def\eea{\end{eqnarray}}

\begin{document}

\title{LOW-ENERGY QCD: CHIRAL COEFFICIENTS, 
$U_A (1)$ BREAKING AND THE QUARK-QUARK INTERACTION}

\author{T.MEISSNER}

\address{Department of Physics and Astronomy, University of South Carolina,\\
Columbia, SC 29208, USA}

\author{M.R.FRANK}

\address{Institute for Nuclear Theory,
University of Washington, \\
Seattle, WA 98195, USA}


\maketitle

\section{A Path from QCD to Chiral
Perturbation Theory}

\begin{table}[b]
\caption{The chiral coefficients using a running coupling
$\alpha (s)$. The parameter choices listed maintain $f_\pi = 86 {\mbox{MeV}}$,
which is the generic scale of dynamical chiral symmetry breaking.
The numbers in parantheses are the ``phenomenological'' values. \label{tab1}}
\vspace{0.4cm}
\begin{center}
\begin{tabular}{|l|l|l|l|l|}
\hline 
\multicolumn{5}{|c|}{$\alpha _1(s)=3\pi s\chi ^2e^{-s/\Delta}/(4\Delta^2)
+\pi d/\mbox{ln}(s/\Lambda^2 +e)$} \\ \hline
$\Delta $ (GeV$^2$) & $\chi $ (GeV) &  
$L_1 $(0.7$\pm $0.5) &  $L_3$(-3.6$\pm $1.3) & $L_5$(1.4$\pm $0.5) 
\\ \hline
0.002 & 1.4 &  0.84 & -4.4 & 1.0  \\ \hline
0.02 & 1.5  &  0.82  &  -4.4  & 1.14    \\ \hline
0.2  & 1.65 &  0.81  &  -4.0  & 1.66    \\ \hline
0.4   & 1.84  &  0.80  &  -3.8  & 2.0   \\  \hline
\end{tabular}
\end{center}
\end{table}

A global color symmetry model (GCM) that is based upon an effective 
quark-quark interaction arises from the QCD partition function
by formally integrating over the gluon fields and truncating the 
expansion in gluon $n$ point functions after $n=2$.\cite{RCP}
This model truncation respects all global symmetries of QCD, 
in particular chiral
symmetry. What is lost is invariance under local color gauge transformations.
Fixing a model form for the gluon propagator $D(s)$, or, what is equivalent, 
the running coupling $\alpha (s)$, specifies a quark-quark interaction.
Furthermore the model allows a $\frac{1}{N_c}$ expansion.
The lowest order (${\cal{O}} (N_c)$) consists in solving the Dyson-Schwinger
equation for the quark self energy including rainbow gluon dressings and the ladder
Bethe-Salpeter equation for mesonic bound states.
One then is able to change the degrees of freedom from quarks
to hadrons by performing an appropriate  variable transformation in 
the generating functional path integral.
For the moment only the Goldstone bosons are taken into account.
We obtain a non-local hadronic interaction between the Goldstone fields,
which can be systematically expanded in the momenta of the Golstone fields and the
current quark mass leading exactly to the form given by 
Gasser and Leutwyler.\cite{gl}
The chiral low energy coefficients $L_i$
are now determined by the dynamics of the quark-quark
interaction and can be numerically calculated from the quark self energy 
functions.\cite{ref1}
As one can see from Tab.1, $L_1$ and $L_3$ are practically independent on the 
quark-quark interaction, whereas $L_5$ depends noticeably on its form.

\section{Triangle Daigrams, $U_A (1)$ Breaking and the $\eta^\prime$ Mass}

Including higher mass mesonic states other than the Goldstone fields
and integrating them out generates triangle diagrams such as 


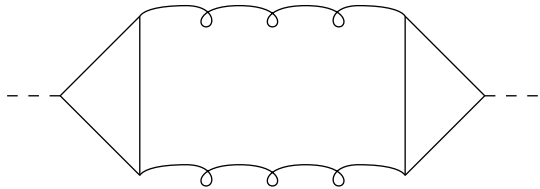
\vspace{0.5cm}
\begin{picture}(250,100)
\Gluon(50,80)(150,80)43
\Gluon(50,20)(150,20)43
\Line(20,50)(50,80)
\Line(20,50)(50,20)
\Line(50,80)(50,20)
\Line(180,50)(150,20)
\Line(180,50)(150,80)
\Line(150,80)(150,20)
\DashLine(0,50)(20,50)5
\DashLine(180,50)(200,50)5
\end{picture}

\vspace{0.5cm}
\noindent Those diagrams are supressed by one order of $\frac{1}{N_c}$ compared
with ones mentioned in the last section.
They break the $U_A (1)$ symmetry and depending on the behavior of the
modelled running coupling $\alpha (s)$ in the infrared can create a mass for
the iso-singlet pseudoscalar Goldstone boson $\eta^\prime$.
This provides a mechanism for an $\eta^\prime$ mass without employing
explicitly topological gauge field configurations (instantons).\cite{ref2}

\end{document}